\documentclass[superscriptaddress,twocolumn,english,floatfix,aps,pra,amsmath,amssymb,longbibliography]
{revtex4-2}
\usepackage[T1]{fontenc}
\usepackage[utf8]{inputenc}
\usepackage{babel}
\usepackage{comment}
\usepackage{color}
\usepackage{times}
\usepackage{epsfig}
\usepackage{subfigure}
\usepackage{amsmath}
\usepackage{amssymb}
\usepackage{graphicx}
\usepackage[export]{adjustbox}
\usepackage[colorlinks=true]{hyperref}
\hypersetup{citecolor=blue,linkcolor=blue,urlcolor=blue}

\begin{document}

\title{Curvature-induced repulsive effect on the lateral Casimir-Polder--van der Waals force}
%

\author{Danilo T. Alves}
\email{danilo@ufpa.br}
\affiliation{Faculdade de F\'{i}sica, Universidade Federal do Par\'{a}, 66075-110, Bel\'{e}m, Par\'{a}, Brazil}
\affiliation{Centro de F\'{i}sica, Universidade do Minho, P-4710-057, Braga, Portugal}

\author{Lucas Queiroz}
\email{lucas.silva@icen.ufpa.br}
\affiliation{Faculdade de F\'{i}sica, Universidade Federal do Par\'{a}, 66075-110, Bel\'{e}m, Par\'{a}, Brazil}
\author{Edson C. M. Nogueira}
\email{edson.moraes.nogueira@icen.ufpa.br}
\affiliation{Faculdade de F\'{i}sica, Universidade Federal do Par\'{a}, 66075-110, Bel\'{e}m, Par\'{a}, Brazil}
\author{N. M. R. Peres}
\email{peres1975@gmail.com}
\affiliation{Centro de F\'{i}sica, Universidade do Minho, P-4710-057, Braga, Portugal}
\affiliation{Departamento de F\'{i}sica, Universidade do Minho, P-4710-057, Braga, Portugal}

\date{\today}

\begin{abstract}
We consider a perfectly conducting infinite cylinder with radius $R$, and investigate the Casimir-Polder (CP) and van der Waals (vdW) interactions with a neutral polarizable particle constrained to move in a plane distant $x_0>R$ from the axis of the cylinder.
We show that when the relative curvature $x_0/R \lesssim 6.44$, this particle, 
under the action of the lateral CP force (which is the projection of the CP force onto the mentioned plane), is attracted to the point on the plane which is closest to the cylinder surface.
On the other hand, when $x_0/R \gtrsim 6.44$, we also show that, for certain particle orientations and anisotropy, the lateral CP force can move the particle away from the cylinder. 
This repulsive behavior of such a component of the CP force reveals a nontrivial dependence of the CP interaction with the surface geometry, specifically of the relative curvature. 
In the vdW regime, we show that a similar nontrivial 
repulsive behavior occurs,
but for the relative curvature $x_0/R \gtrsim 2.18$, which means that this effect requires 
a smaller cylinder curvature in the vdW regime than in the CP one.
In addition, we also show that there are classical counterparts of these effects, involving a neutral particle with a permanent electric dipole moment.
The prediction of such geometric effects on this force may be relevant for a better controlling of the interaction between a particle and a curved surface in classical and quantum physics.
%
\end{abstract}

\maketitle
	
\section{Introduction}

The quantum electromagnetic dispersive force between a polarizable particle and a macroscopic body is, generally, denominated as the Casimir-Polder (CP) force, or as van der Waals (vdW) force in the nonretarded regime \cite{Eisenschitzand-London-ZeitPhys-1930,London-ZeitPhys-1930,Casimir-Polder-Nature-1946,Casimir-Polder-PhysRev-1948, Feinberg-Sucher-PRA-1970}.
Their attractive or repulsive character can be influenced, for instance, by the particle anisotropy and geometry of the bodies \cite{Levin-PRL-2010, Abrantes-PRA-2018,Eberlein-PRA-2011,Marchetta-PRA-2021,Buhmann-IJMPA-2016}.
For example, an anisotropic polarizable particle can feel a normal repulsive force when it is put on the symmetry axis of a thin metal plate with a hole \cite{Levin-PRL-2010} or of a perfectly conducting toroid \cite{Abrantes-PRA-2018}. 
When corrugations are considered to the body surface, lateral CP-vdW forces appear, and nontrivial geometric effects can be predicted, especially for these particular components of these forces \cite{Dalvit-PRL-2008, Dalvit-JPA-2008,Dobrich-PRD-2008,Messina-PRA-2009, Moreno-NJP-2010, Reyes-PRA-2010, Moreno-PRL-2010, Bimonte-PRD-2014, Marachevsky-TMP-2015, Bennett-PRA-2015, Buhmann-IJMPA-2016, Nogueira-PRA-2021,Queiroz-PRA-2021,Nogueira-PRA-2022}. 
For instance, in Ref. \cite{Nogueira-PRA-2021}, it was shown that, when a sinusoidal corrugation with period $L$ is introduced on an infinite flat conducting surface at $x=0$, an anisotropic particle, kept constrained to move on a plane $x=x_0$ above the surface, feels a lateral vdW force that leads it to the nearest corrugation peak 
when sufficiently small values of the ratio $x_0/L$ are considered.
Otherwise, increasing the ratio $x_0/L$, 
the lateral vdW force can change its sign, moving the particle away from the corrugation peak \cite{Nogueira-PRA-2021}.
Moreover, in Ref. \cite{Nogueira-PRA-2022}, the collective effect of the peaks was removed by considering a flat surface, still infinite, with only a single slight protuberance.
In this case, it was shown that a sign inversion in the lateral vdW force can still occur when the ratio $x_0/l$ increases, with $l$ being the characteristic width of the protuberance.
In addition, in Ref. \cite{Nogueira-PRA-2022}, it was observed a clear distinction between sign inversions in the lateral vdW force originated by a purely individual effect from those originated only in the presence of a collective of protuberances.

\begin{figure}
\centering
\epsfig{file=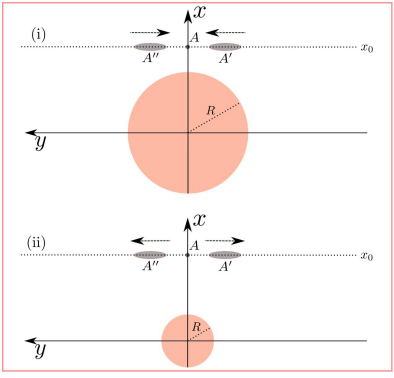, width=0.6 \linewidth}
\caption{
Illustration of our initial conjecture, which is investigated (and confirmed) in the present paper: the sign inversion in the lateral force for an infinite cylinder of radius $R$, in both retarded and nonretarded regimes, as the relative curvature increases.
The arrows represent the lateral force acting on a neutral anisotropic polarizable particle (elliptic figures), kept constrained to move on a given plane $x=x_0$ (horizontal dotted lines).
In (i), the particle feels a lateral force $F_{\text{CP-vdW}}^{(y)}$ that takes it back to $A$, which is the point on the plane closest to the cylinder surface. 
In (ii), for an increased relative curvature $x_0/R$, $F_{\text{CP-vdW}}^{(y)}$ moves the particle away from $A$, or from the cylinder. 
This repulsive behavior reveals a nontrivial dependence of the CP-vdW interaction on the cylinder relative curvature.
}
\label{fig:introducao-cilindro}
\end{figure}

In the present paper, motivated by the predictions from Refs. \cite{Nogueira-PRA-2021,Nogueira-PRA-2022}, we investigate the occurrence of a similar sign inversion in the lateral force for a perfectly reflecting infinite cylinder, in both retarded (CP) and nonretarded (vdW) regimes. 
Our initial conjecture (which will be confirmed along this paper), 
is that a neutral anisotropic polarizable particle, kept constrained to move on a given plane distant $x_0>R$ from the cylinder axis (with $R$ being the cylinder radius), could feel a lateral CP-vdW force that takes it back to the point closest to the cylinder when it is slightly dislocated from this point [see Fig. \ref{fig:introducao-cilindro}(i)].
On the other hand, for sufficiently large values of the relative curvature $x_0/R$, the lateral CP-vdW force would change its sign, moving the particle away from the cylinder [see Fig. \ref{fig:introducao-cilindro}(ii)].

The cylindrical geometry has been considered in problems involving dispersive interactions \cite{
Rosenfeld-1974,GU-JCIS-1999,Kirsch-ACIS-2003,Baglov-PRB-2005, Bordag-PRD-2006,Gies-PRL-2006,Eberlein-PRA-2007,Eberlein-PRA-2009,Bezerra-EPJC-2011,Valchev2017-AIP-Conf-Proceed}, 
with curvature effects discussed, for example, in Refs. \cite{Gies-PRL-2006,Valchev2017-AIP-Conf-Proceed}.
In addition, the problem of how to probe geometric effects on the quantum vacuum fluctuations,
by considering the lateral CP force, has been addressed \cite{Dalvit-PRL-2008,Dalvit-JPA-2008}.
Thus, the investigation carried out here, by predicting 
a curvature-induced repulsive effect on the lateral CP-vdW force,
may be relevant for a better controlling of the interaction between a particle and a curved surface, as well as
an additional way to understand geometric effects on the quantum vacuum fluctuations.

The paper is organized as follows. 
In Sec. \ref{sec:CP-vdW}, we investigate the CP-vdW interaction
between a perfectly reflecting infinite cylinder and a neutral polarizable particle,
kept constrained to move on a given plane near the cylinder.
We also discuss the classical interaction involving a neutral particle with a permanent electric dipole moment.
In Sec. \ref{sec:final} we discuss some implications of our results,
and present our final comments.

\section{Casimir-Polder--van der Waals interaction}
\label{sec:CP-vdW}

\begin{figure}
\centering
\epsfig{file=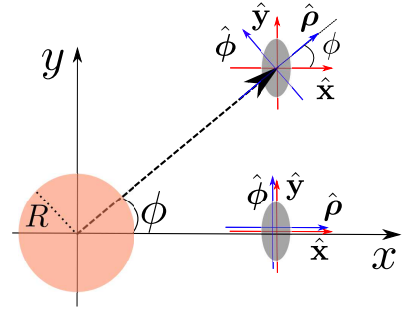,  width=0.6 \linewidth}
\caption{Illustration of a conducting cylinder, a neutral polarizable particle (shown in two positions), and the Cartesian system $xyz$ ($z$ axis is perpendicular to the paper plane). 
The particle is oriented in space in  such a way that its principal axes are parallel to the $xyz$ axes. 
In one position of the particle, for which $\phi=0$, these principal axes coincide with the cylindrical ones. 
When the particle is translated to another position for which $\phi\neq 0$, these axes do not coincide anymore.}
\label{fig:particula-cilindro}
\end{figure}

Let us start by considering the retarded (CP) regime, and investigate the quantum energy interaction $U_{\text{CP}}$ between a perfectly reflecting infinite cylinder and a neutral polarizable particle, oriented in space in such a way that its principal axes are parallel to the $xyz$ axes of a Cartesian system, as illustrated in Fig. \ref{fig:particula-cilindro}.
In this way, an electric field applied on the particle, along any one of these Cartesian axes, induces a dipole moment in the same direction \cite{Feynman-Lectures-vol-2} (this is just a convenient choice that does not imply any loss of generality for the conclusions obtained in the present work).
We also consider this particle characterized by a frequency dependent polarizability tensor $\overleftrightarrow{\alpha}\left(\omega\right)$, whose representation $\boldsymbol{\alpha}^{(\text{cart})}\left(\omega\right)$ in the Cartesian system is given by 
\begin{equation}
	\boldsymbol{\alpha}^{(\text{cart})}\left(\omega\right)=\left[\begin{array}{ccc}
		\alpha_{xx}\left(\omega\right) & 0 & 0\\
		0 & \alpha_{yy}\left(\omega\right) & 0\\
		0 & 0 & \alpha_{zz}\left(\omega\right)
	\end{array}\right].
\end{equation}
The tensor $\overleftrightarrow{\alpha}\left(\omega\right)$, for a translated particle as illustrated in  Fig. \ref{fig:particula-cilindro}, is represented in cylindrical coordinate system $(\rho,\phi,z)$ by means of the matrix $\boldsymbol{\alpha}^{(\text{cil})}\left(\omega,\phi\right)$ given by
\begin{equation}
	\boldsymbol{\alpha}^{(\text{cil})}\left(\omega,\phi\right)=\left[\begin{array}{ccc}
		\alpha_{\rho\rho}\left(\omega,\phi\right) & \alpha_{\rho\phi}\left(\omega,\phi\right) & 0\\
		\alpha_{\phi\rho}\left(\omega,\phi\right) & \alpha_{\phi\phi}\left(\omega,\phi\right) & 0\\
		0 & 0 & \alpha_{zz}\left(\omega\right)
	\end{array}\right],
	\label{eq:alpha-primo-phi}
\end{equation}
where $\alpha_{\rho\rho}\left(\omega,\phi\right)$
$=\alpha_{yy}\left(\omega\right)\sin^{2}\phi+\alpha_{xx}\left(\omega\right)\cos^{2}\phi$,
$\alpha_{\rho\phi}\left(\omega,\phi\right)$$=\alpha_{\phi\rho}\left(\omega,\phi\right)=\left[\alpha_{yy}\left(\omega\right)-\alpha_{xx}\left(\omega\right)\right]\sin\phi\cos\phi$,
$\alpha_{\phi\phi}\left(\omega,\phi\right)$
$=\alpha_{yy}\left(\omega\right)\cos^{2}\phi+\alpha_{xx}\left(\omega\right)\sin^{2}\phi$.
Note that $	\boldsymbol{\alpha}^{(\text{cart})}\left(\omega\right)=
\boldsymbol{\alpha}^{(\text{cil})}\left(\omega,0\right)$, $\cos\phi=x/\rho$, and
 $\sin\phi=y/\rho$.

To investigate the behavior of $U_{\text{CP}}$, 
we take into account Eqs. (17) and (27)-(29) found in Ref. \cite{Eberlein-PRA-2009} 
(the same formulas are also obtained in Ref. \cite{Bezerra-EPJC-2011}).
Specifically, we consider in these formulas the limit $\lambda_{ji}\to 0$
($E_{ji}\to \infty$), with $\lambda_{ji}$ and $E_{ji}$ being, respectively, 
the wavelength and energy of a typical transition between the states
$i$ and $j$ of the particle  \cite{Eberlein-PRA-2009}.
The obtained formula can be written
in terms of the static polarizability tensor $\overleftrightarrow{\alpha}\left(0\right)$,
specifically of the diagonal components $\rho\rho$, $\phi\phi$, and $zz$. 
For a translated particle (see Fig. \ref{fig:particula-cilindro}),
these components depend on $\phi$ according to Eq. \eqref{eq:alpha-primo-phi}, so that
we use $\alpha_{\rho\rho}(0,\phi)$
$=\alpha_{xx}\left(0\right)x^{2}/\rho^2+\alpha_{yy}\left(0\right)y^{2}/\rho^2$,
and $\alpha_{\phi\phi}(0,\phi)$
$=\alpha_{xx}\left(0\right)y^{2}/\rho^2+\alpha_{yy}\left(0\right)x^{2}/\rho^2$ 
in the obtained formula. Thus, one has
\begin{align}
	U_{\text{CP}}(\overline{x},\overline{y})&=-\frac{1}{\left(4\pi\right)^{2}\epsilon_{0}R^{4}}\nonumber\\
	&\times\bigg[\Xi_{\rho}^{(\text{CP})}\left(\overline{\rho}\right)\left(\alpha_{xx}\left(0\right)\frac{\overline{x}^{2}}{\overline{\rho}^{2}}
	+\alpha_{yy}\left(0\right)\frac{\overline{y}^{2}}{\overline{\rho}^{2}}\right)\nonumber\\
	&+\Xi_{\phi}^{(\text{CP})}\left(\overline{\rho}\right)\left(\alpha_{xx}\left(0\right)\frac{\overline{y}^{2}}{\overline{\rho}^{2}}
	+\alpha_{yy}\left(0\right)\frac{\overline{x}^{2}}{\overline{\rho}^{2}}\right)\nonumber\\
	&+\Xi_{z}^{(\text{CP})}\left(\overline{\rho}\right)
	\alpha_{zz}\left(0\right)\bigg],
	\label{eq:Eberlein-Zietal-CP}
\end{align}
where:
\begin{align}
	\Xi_{\rho}^{(\text{CP})}\left(\overline{\rho}\right)&=
	2\sum_{m=0}^{\infty}{\vphantom{\sum}}'\int_{0}^{\infty}duu
	\left\{u^{2}\frac{I_{m}\left(u\right)}{K_{m}\left(u\right)}\left[K_{m}^{\prime}\left(u\overline{\rho}\right)\right]^{2}\right.
	\nonumber
	\\
	&\left.-\frac{m^{2}}{\overline{\rho}^{2}}\frac{I_{m}^{\prime}\left(u\right)}{K_{m}^{\prime}\left(u\right)}\left[K_{m}\left(u\overline{\rho}\right)\right]^{2}\right\},
	\label{eq:Xi-rho}
\end{align}
\begin{align}
	\Xi_{\phi}^{(\text{CP})}\left(\overline{\rho}\right)&=
	2\sum_{m=0}^{\infty}{\vphantom{\sum}}'\int_{0}^{\infty}duu\left\{-u^{2}\frac{I_{m}^{\prime}\left(u\right)}{K_{m}^{\prime}\left(u\right)}\left[K_{m}^{\prime}\left(u\overline{\rho}\right)\right]^{2}\right.
	\nonumber\\
	&\left.+\frac{m^{2}}{\overline{\rho}^{2}}\frac{I_{m}\left(u\right)}{K_{m}\left(u\right)}\left[K_{m}\left(u\overline{\rho}\right)\right]^{2}\right\},
	\label{eq:Xi-phi}
\end{align}
\begin{equation}
	\Xi_{z}^{(\text{CP})}\left(\overline{\rho}\right)=
	4\sum_{m=0}^{\infty}{\vphantom{\sum}}'\int_{0}^{\infty}duu^{3}\frac{I_{m}\left(u\right)}{K_{m}\left(u\right)}\left[K_{m}\left(u\overline{\rho}\right)\right]^{2},
	\label{eq:Xi-z}
\end{equation}
with $\sum_{m=0}^{\infty}{\vphantom{\sum}}'f_m=\frac{1}{2}f_0+\sum_{m=1}^{\infty}f_m$,
$R$ being the cylinder radius, $\overline{x}={x}/{R}$, $\overline{y}={y}/{R}$, $\overline{\rho}={\rho}/{R}$.

\begin{figure}
\centering
\epsfig{file=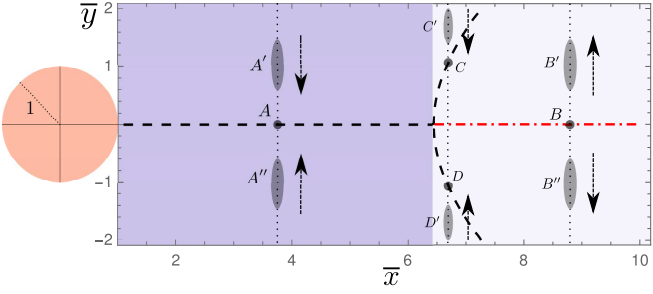,  width=1 \linewidth}
\caption{
Some features of the CP interaction between a perfectly reflecting cylinder and a polarizable particle (illustrated in several positions by the ellipsoidal figures), with $\alpha_{xx}(0)=\alpha_{zz}(0)=0$, and kept constrained to move on a given plane  $\overline{x}=\overline{x}_0>1$ (three of them represented by the vertical dotted lines).
Note that the cylinder circular section and the axes $\overline{x}$ and $\overline{y}$ are represented at a same scale. 
We also remark that the dashed and dot-dashed lines are plotted taking into account Eq. \eqref{eq:Eberlein-Zietal-CP}.
We have $\partial U_{\text{CP}}/\partial y=0$ on both dashed and dot-dashed lines, with the dashed line corresponding to minimum points of $U_{\text{CP}}(x_0,y)$, whereas the dot-dashed one corresponding to maximum points.
In the dark region ($1 < \overline{x} < 6.44$), when the particle is dislocated, along the $y$-axis, for instance, from the point $A$ to $A^\prime$ (or $A^{\prime\prime}$), it feels a force $F_{\text{CP}}^{(y)}$ (represented by the arrows) which takes it back to $A$. 
In the light region ($6.44 < \overline{x}$), when the particle is dislocated, for instance, from the point $B$ to $B^\prime$ (or $B^{\prime\prime}$), it feels a force $F_{\text{CP}}^{(y)}$ which moves it away from $B$, and, consequently, away from the cylinder.
This sign inversion in $F_{\text{CP}}^{(y)}$
is a nontrivial geometric effect regulated by the relative curvature $\overline{x}=x/R$.
When the particle is dislocated, along $\overline{x}=\overline{x}_0$, for instance, from the point $C$ to $C^\prime$ (or from $D$ to $D^{\prime}$), it feels a force $F_{\text{CP}}^{(y)}$ that moves it back to $C$ ($D$).
}
\label{fig:brasilia3}
\end{figure}

\begin{figure}[h]
\centering  
\subfigure[]{\label{fig:CP6}\epsfig{file=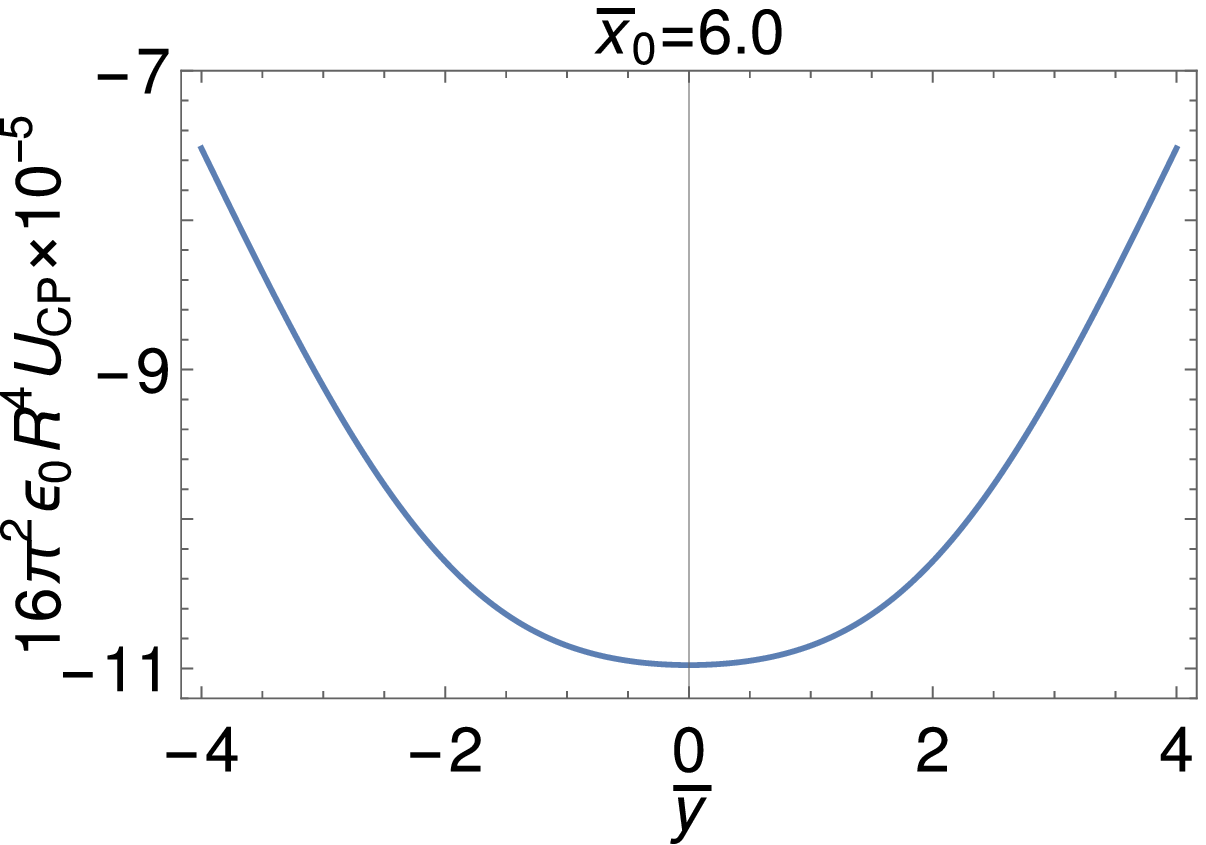, width=0.48 \linewidth}}
\hspace{2mm}
\subfigure[]{\label{fig:CP644}\epsfig{file=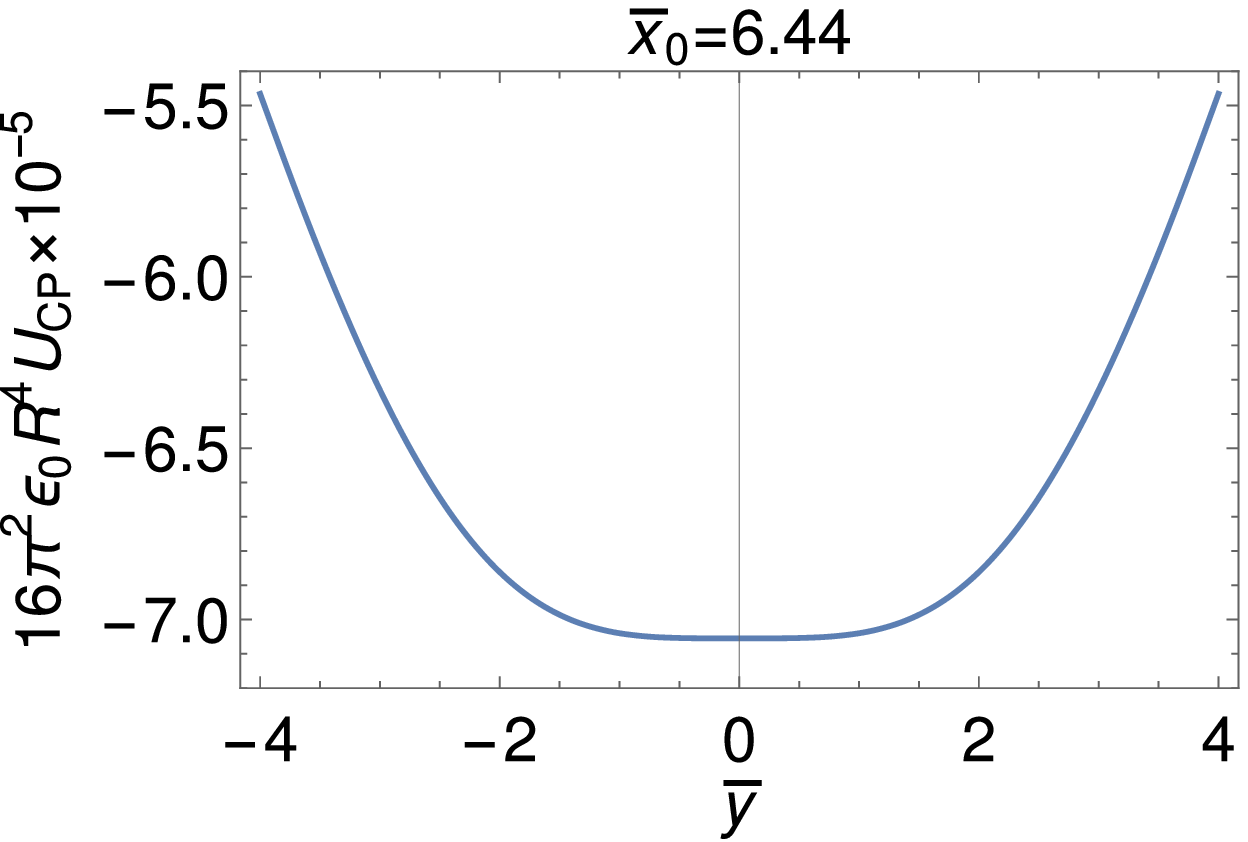, width=0.48 \linewidth}}
\hspace{2mm}
\subfigure[]{\label{fig:CP7}\epsfig{file=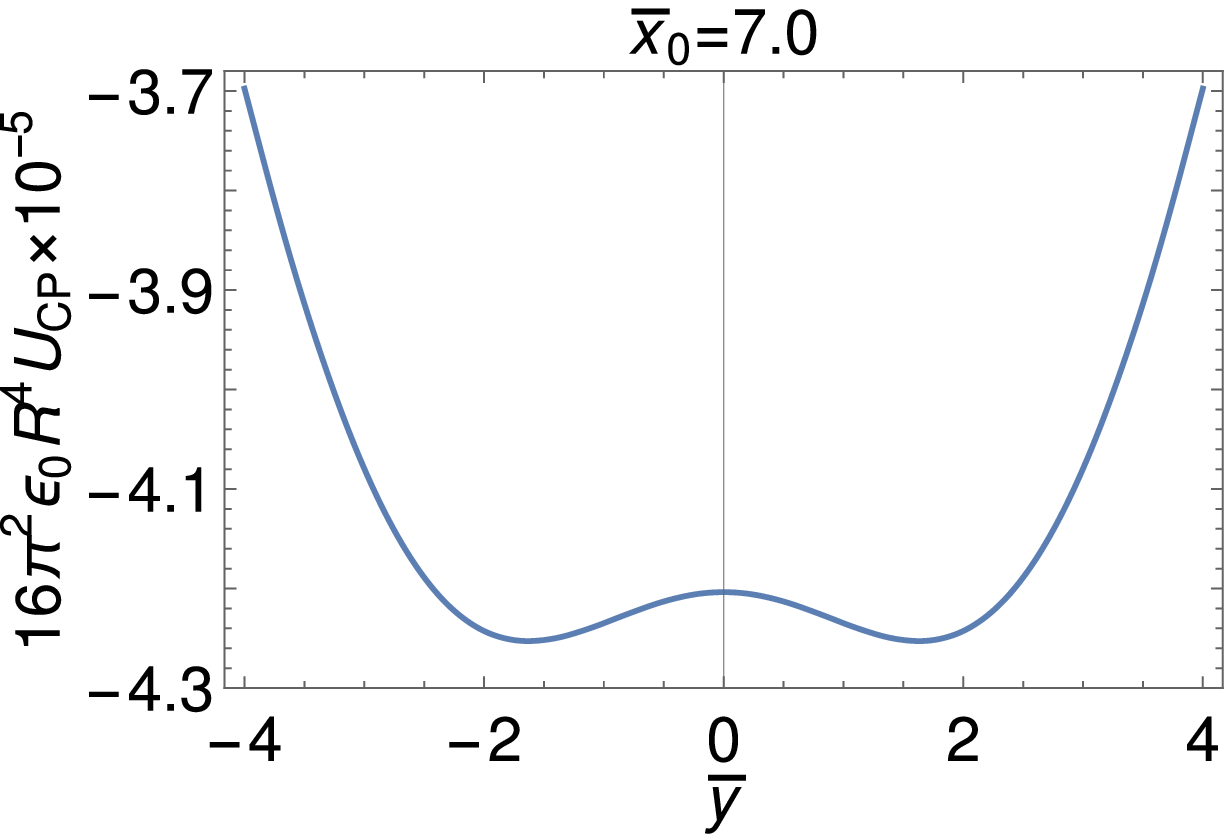, width=0.48 \linewidth}}
\hspace{2mm}
\subfigure[]{\label{fig:CP75}\epsfig{file=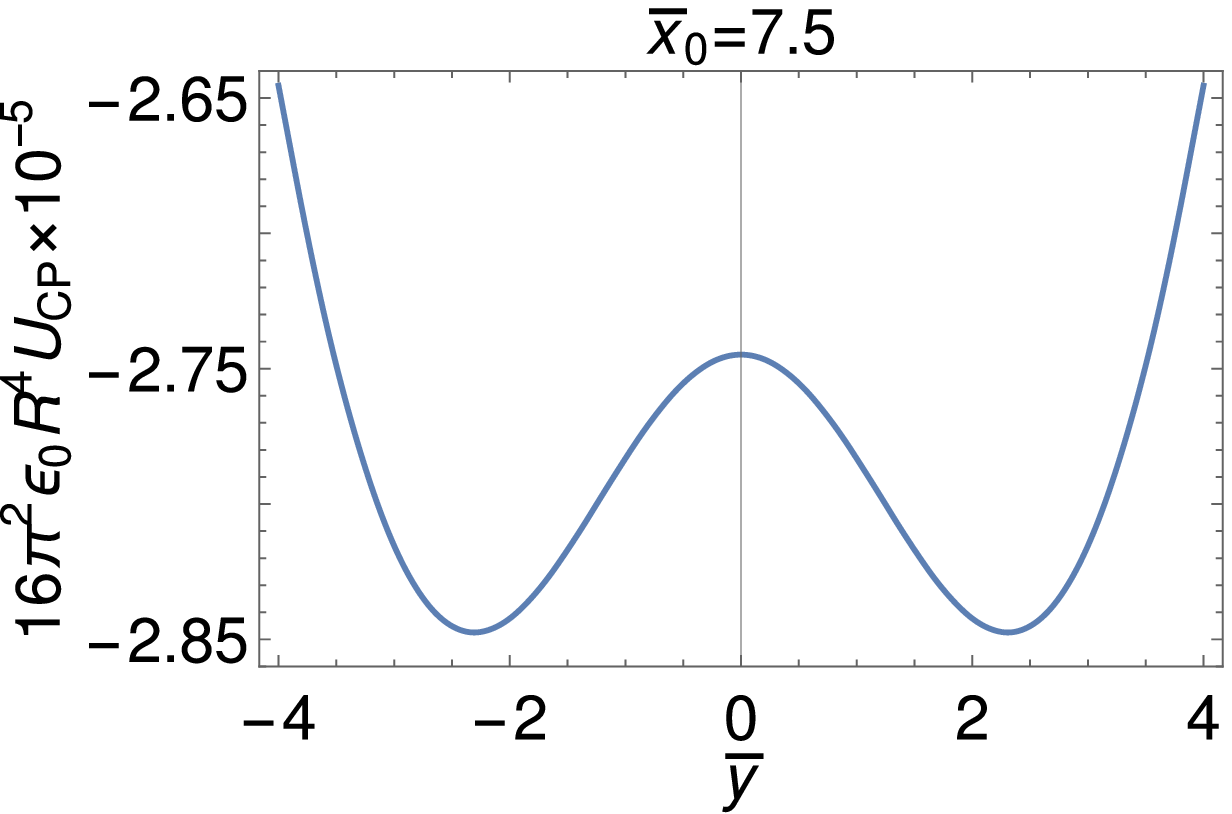, width=0.48 \linewidth}}
\caption{
The behavior of $U_{\text{CP}}(\overline{x}_0,\overline{y})$
for some values of $\overline{x}_0$.
In (a), $\overline{x}_0=6.00$. 
In (b), $\overline{x}_0=6.44$.
In (c), $\overline{x}_0=7.00$. 
In (d), $\overline{x}_0=7.50$.
Panels (a) and (b) correspond to the dark region in Fig. \ref{fig:brasilia3}, whereas (c) and (d) to the light one.
}
\label{fig:CP}
\end{figure}

We start our analysis with the idealized case in which $\alpha_{xx}(0)= \alpha_{zz}(0)= 0$.
The behavior of $U_{\text{CP}}(x,y)$ is
such that the CP force ${\bf F}_{\text{CP}}=F_{\text{CP}}^{(x)}{\bf \hat{x}}+F_{\text{CP}}^{(y)}{\bf \hat{y}=}-\boldsymbol{\nabla}U_{\text{CP}}$ always attracts the particle to the cylinder when no constraint is imposed to the particle.
Now, let us keep the particle constrained to move on a given plane  $\overline{x}=\overline{x}_0>1$.
From Eq. \eqref{eq:Eberlein-Zietal-CP}, we have that
$\partial U_{\text{CP}}/\partial y=0$ along the dashed and dot-dashed lines shown in Fig. \ref{fig:brasilia3}, with $[\partial^{2}U_{\text{CP}}(x_0,y)/\partial y^{2}]>0$ [minimum points of $U_{\text{CP}}(x_0,y)$] along the dashed line, and $[\partial^{2}U_{\text{CP}}(x_0,y)/\partial y^{2}]<0$ [maximum points of $U_{\text{CP}}(x_0,y)$] along the dot-dashed one.
Thus, in Fig. \ref{fig:brasilia3}, a particle in the dark region ($1 < \overline{x} < 6.44$), slightly dislocated from a point in the plane $y=0$, 
feels a force $F_{\text{vdW}}^{(y)}$ that takes it back to $y=0$. 
On the other hand, in the light region ($6.44 < \overline{x}$), the particle
is moved away from $y=0$, 
or, in other words, away from the cylinder (see Fig. \ref{fig:brasilia3}).
This sign inversion in $F_{\text{CP}}^{(y)}$,
changing from an attractive character to a repulsive one,
is a nontrivial geometric effect regulated by the relative curvature $\overline{x}$.
The behavior of ${U}_{\text{CP}}(\overline{x}_0,\overline{y})$, for some values of $\overline{x}_0$,
is shown in Fig. \ref{fig:CP}.

\begin{figure}
\epsfig{file=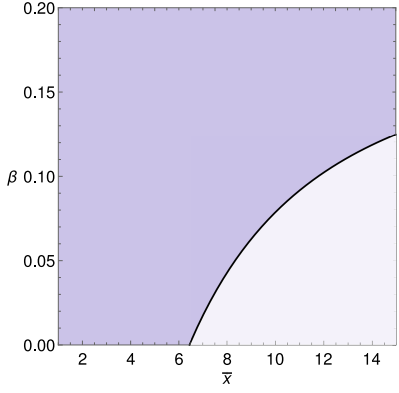,  width=0.6 \linewidth}
\caption{
For a particle characterized by $\alpha_{xx}(0)=\alpha_{zz}(0)=\beta \alpha_{yy}(0)$, it is shown the configurations of $\beta$ and $\overline{x}$ for which $y=0$ is a minimum (dark region) or a maximum (light region) point of $U_{\text{CP}}(x_0,y)$.
The border between these two regions is given by the curve 
$\overline{x}=\gamma(\beta)$.
We remark that $\gamma(0)\approx 6.44$ (the situation discussed in Fig. \ref{fig:brasilia3}), and that above $\beta\approx 0.2$ we just have a dark region.
}
\label{fig:regioes-cp}
\end{figure}

For a particle with $\alpha_{xx}(0)=\alpha_{zz}(0)=\beta \alpha_{yy}(0)$, with $0\leq \beta < 0.2$, we have a behavior similar to that shown in Fig. \ref{fig:brasilia3}, but with the border between the dark and light regions occurring at a value $\overline{x}=\gamma(\beta)$.
In Fig. \ref{fig:regioes-cp}, we show the configurations of $\beta$ and $\overline{x}$ for which $y=0$ is a minimum (dark region) or a maximum (light region) point of $U_{\text{CP}}(x_0,y)$.
Thus, the dark region of this figure represents configurations for which $[\partial^{2}U_{\text{CP}}(x_0,y)/\partial y^{2}]_{y=0}>0$, and the behavior of the lateral force is similar to that described for the dark region in Fig. \ref{fig:brasilia3}.
Besides this, the light region represents $[\partial^{2}U_{\text{CP}}(x_0,y)/\partial y^{2}]_{y=0}<0$, and the behavior is similar to that described for the light region in Fig. \ref{fig:brasilia3}.
The border between the two regions in Fig. \ref{fig:regioes-cp} is given by the curve 
$\overline{x}=\gamma(\beta)$.
Note that, $\gamma(0)\approx 6.44$, which corresponds to the situation discussed in Fig. \ref{fig:brasilia3}.
We remark that for $\beta \gtrsim 0.2$ we just have a dark region, so that any repulsive effect of the lateral force is suppressed.

\begin{figure}
\epsfig{file=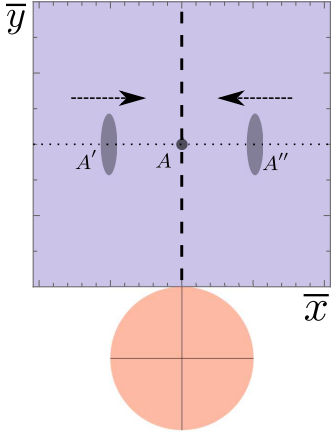,width=0.5 \linewidth}
\caption{
Illustration of some features of the vdW-CP interaction between a perfectly reflecting cylinder and a polarizable particle (represented in two positions by the ellipsoidal figures), kept constrained to move on a given plane  $\overline{y}=\overline{y}_0>1$ (represented by the dotted line).
The dashed line corresponds to minimum points of $U_{\text{CP-vdW}}(x,y_0)$, so that when the particle is dislocated, 
for instance, from the point $A$ to $A^\prime$ (or $A^{\prime\prime}$), it feels a force $F_{\text{CP-vdW}}^{(x)}$ (represented by the arrows) which takes it back to $A$.
}
\label{fig:brasilia-sem-asa}
\end{figure}

It is worth to mention that these curvature-induced effects are affected by the particle orientation in relation to the plane where it is kept constrained to move.
As an example, let us consider the particle with $\alpha_{xx}(0)=\alpha_{zz}(0)=\beta \alpha_{yy}(0)$ ($0 \leq \beta \leq 1$) oriented as illustrated in Fig. \ref{fig:particula-cilindro}, and kept constrained to move on a given plane $y=y_0>1$ (see Fig. \ref{fig:brasilia-sem-asa}).
In this case, we have that the point $x=0$ is always a minimum point of $U_{\text{CP}}(x,y_0)$, independent on the value of $\beta$ or $\overline{y}$.
This means that, when this particle is slightly dislocated from a point in the plane $x=0$, it feels a force $F_{\text{vdW}}^{(x)}$ that takes it back to $x=0$ in all values of $\overline{y}$ (see Fig. \ref{fig:brasilia-sem-asa}).

Now, let us consider the nonretarded (vdW) regime, and investigate the quantum energy interaction $U_{\text{vdW}}$. 
To this end, we take the limit $\lambda_{ji}\to \infty$ ($E_{ji}\to 0$) in Eqs. (17) and (27)-(29) found in Ref. \cite{Eberlein-PRA-2009}. 
After this, one obtains the formula for
$U_{\text{vdW}}$ given by Eq. (19) in Ref. \cite{Eberlein-PRA-2007}, 
which is written in terms of the expectation values $\langle \hat{d}_{\rho}^2\rangle$,
$\langle \hat{d}_{\phi}^2\rangle$, and $\langle \hat{d}_{z}^2\rangle$,
where $\hat{d}_s$ are components of the dipole moment operator.
For a translated particle (see Fig. \ref{fig:particula-cilindro}),
the components $\alpha_{\rho\rho}$ and  $\alpha_{\phi\phi}$ depend on $\phi$
according to Eq. \eqref{eq:alpha-primo-phi}, so that
we write $\langle \hat{d}_{\rho}^2\rangle(\phi)= \frac{\hbar}{\pi}\int_{0}^{\infty}d\xi\,\alpha_{\rho\rho} \left(i\xi,\phi\right)$
$=\langle \hat{d}_{x}^2\rangle x^{2}/\rho^2+\langle \hat{d}_{y}^2\rangle y^{2}/\rho^2$, and
$\langle \hat{d}_{\phi}^2\rangle(\phi)= \frac{\hbar}{\pi}\int_{0}^{\infty}d\xi\,\alpha_{\phi\phi} \left(i\xi,\phi\right)$
$=\langle \hat{d}_{x}^2\rangle y^{2}/\rho^2+\langle \hat{d}_{y}^2\rangle x^{2}/\rho^2$.
Thus, one obtains
\begin{align}
U_{\text{vdW}}(\overline{x},\overline{y})&=-\frac{1}{4\pi\epsilon_{0}R^{3}}\nonumber\\
&\times\bigg[\Xi_{\rho}^{(\text{vdW})}\left(\overline{\rho}\right)\left(\langle\hat{d}_{x}^{2}\rangle\frac{\overline{x}^{2}}{\overline{\rho}^{2}}+\langle\hat{d}_{y}^{2}\rangle\frac{\overline{y}^{2}}{\overline{\rho}^{2}}\right)
\nonumber\\&+\Xi_{\phi}^{(\text{vdW})}\left(\overline{\rho}\right)\left(\langle\hat{d}_{x}^{2}\rangle\frac{\overline{y}^{2}}{\overline{\rho}^{2}}+\langle\hat{d}_{y}^{2}\rangle\frac{\overline{x}^{2}}{\overline{\rho}^{2}}\right)
\nonumber\\&+\Xi_{z}^{(\text{vdW})}\left(\overline{\rho}\right)\langle\hat{d}_{z}^{2}\rangle\bigg],
\label{eq:Eberlein-Zietal-vdW}
\end{align}
where:
\begin{equation}
\Xi_{\rho}^{(\text{vdW})}\left(\overline{\rho}\right)=\frac{2}{\pi}\sum_{m=0}^{\infty}{\vphantom{\sum}}'\int_{0}^{\infty}duu^{2}\frac{I_{m}\left(u\right)}{K_{m}\left(u\right)}\left[K_{m}^{\prime}\left(u\overline{\rho}\right)\right]^{2},
\end{equation}
\begin{equation}
\Xi_{\phi}^{(\text{vdW})}\left(\overline{\rho}\right)=\frac{2}{\pi}\sum_{m=1}^{\infty}\frac{m^{2}}{\overline{\rho}^{2}}\int_{0}^{\infty}du\frac{I_{m}\left(u\right)}{K_{m}\left(u\right)}\left[K_{m}\left(u\overline{\rho}\right)\right]^{2},
\end{equation}
\begin{equation}
\Xi_{z}^{(\text{vdW})}\left(\overline{\rho}\right)=\frac{2}{\pi}\sum_{m=0}^{\infty}{\vphantom{\sum}}'\int_{0}^{\infty}duu^{2}\frac{I_{m}\left(u\right)}{K_{m}\left(u\right)}\left[K_{m}\left(u\overline{\rho}\right)\right]^{2},
\end{equation}
with $\langle \hat{d}_{j}^2\rangle = \frac{\hbar}{\pi}\int_{0}^{\infty}d\xi\,\alpha_{jj}\left(i\xi\right)$,
for $j=x,y,z$.

\begin{figure}
\centering
\epsfig{file=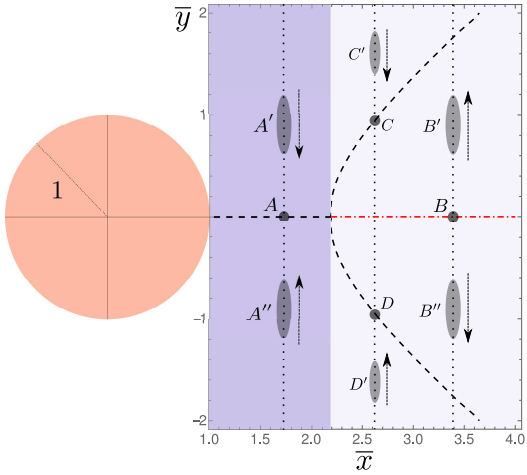,  width=0.8 \linewidth}
\caption{
Some features of the vdW interaction between a perfectly reflecting cylinder and a polarizable particle (represented in several positions by the ellipsoidal figures), with
$\langle \hat{d}_{x}^2\rangle = \langle \hat{d}_{z}^2\rangle = 0$, and kept constrained to move on a given plane  $\overline{x}=\overline{x}_0>1$ (three of them represented by the vertical dotted lines).
Note that the cylinder circular section and the axes $\overline{x}$ and $\overline{y}$ are represented at a same scale. 
We also remark that the dashed and dot-dashed lines are plotted taking into account Eq. \eqref{eq:Eberlein-Zietal-vdW}.
We have $\partial U_{\text{vdW}}/\partial y=0$ on both dashed and dot-dashed lines, with the dashed line corresponding to minimum points of $U_{\text{vdW}}(x_0,y)$, whereas the dot-dashed one corresponding to maximum points.
In dark region ($1 < \overline{x} < 2.18$), when the particle is dislocated, along the $y$-axis, for instance, from the point $A$ to $A^\prime$ (or $A^{\prime\prime}$), it feels a force $F_{\text{vdW}}^{(y)}$ (represented by the arrows) which takes it back to $A$. 
In light region ($2.18 < \overline{x}$), when the particle is dislocated, for instance, from the point $B$ to $B^\prime$ (or $B^{\prime\prime}$), it feels a force $F_{\text{vdW}}^{(y)}$ which moves it away from $B$, and, consequently, away from the cylinder.
This sign inversion in $F_{\text{vdW}}^{(y)}$ is a nontrivial geometric effect regulated by the relative curvature $\overline{x}$.
When the particle is dislocated, along $\overline{x}=\overline{x}_0$, for instance, from the point $C$ to $C^\prime$ (or from $D$ to $D^{\prime}$), it feels a force $F_{\text{vdW}}^{(y)}$ that moves it back to $C$ ($D$).
}
\label{fig:brasilia2}
\end{figure}

\begin{figure}[h]
\centering  
\subfigure[]{\label{fig:vdW15}\epsfig{file=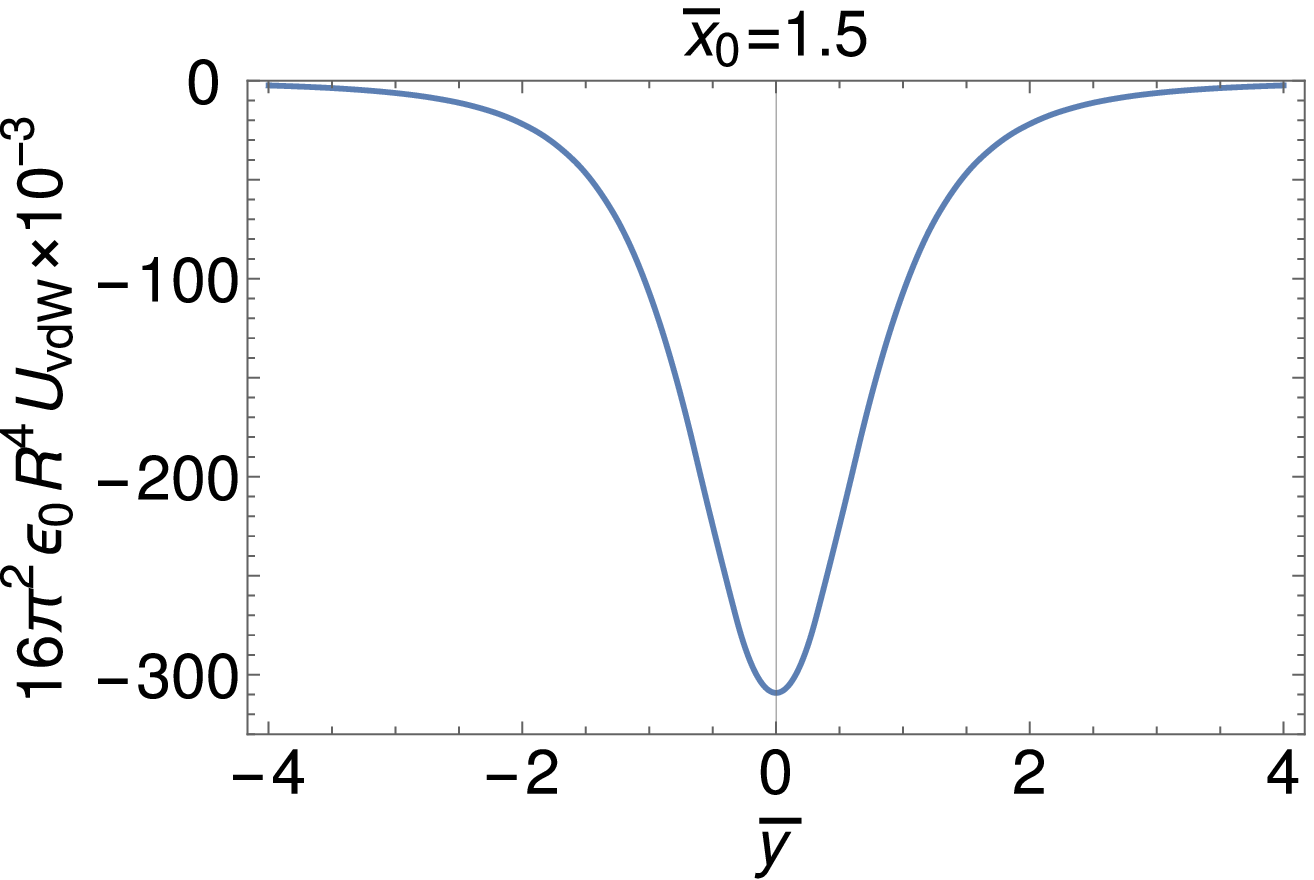, width=0.48 \linewidth}}
\hspace{2mm}
\subfigure[]{\label{fig:vdW218}\epsfig{file=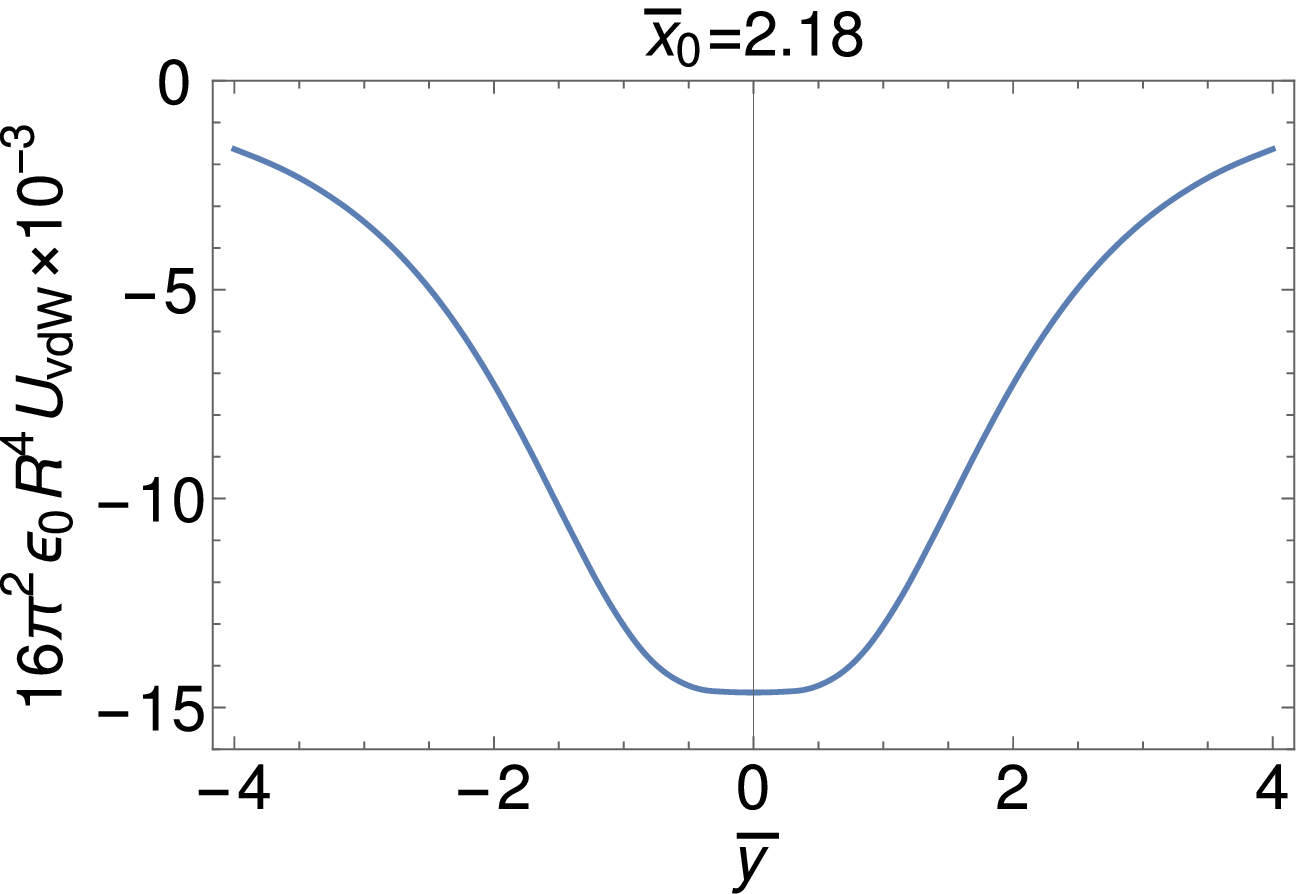, width=0.48 \linewidth}}
\hspace{2mm}
\subfigure[]{\label{fig:vdW25}\epsfig{file=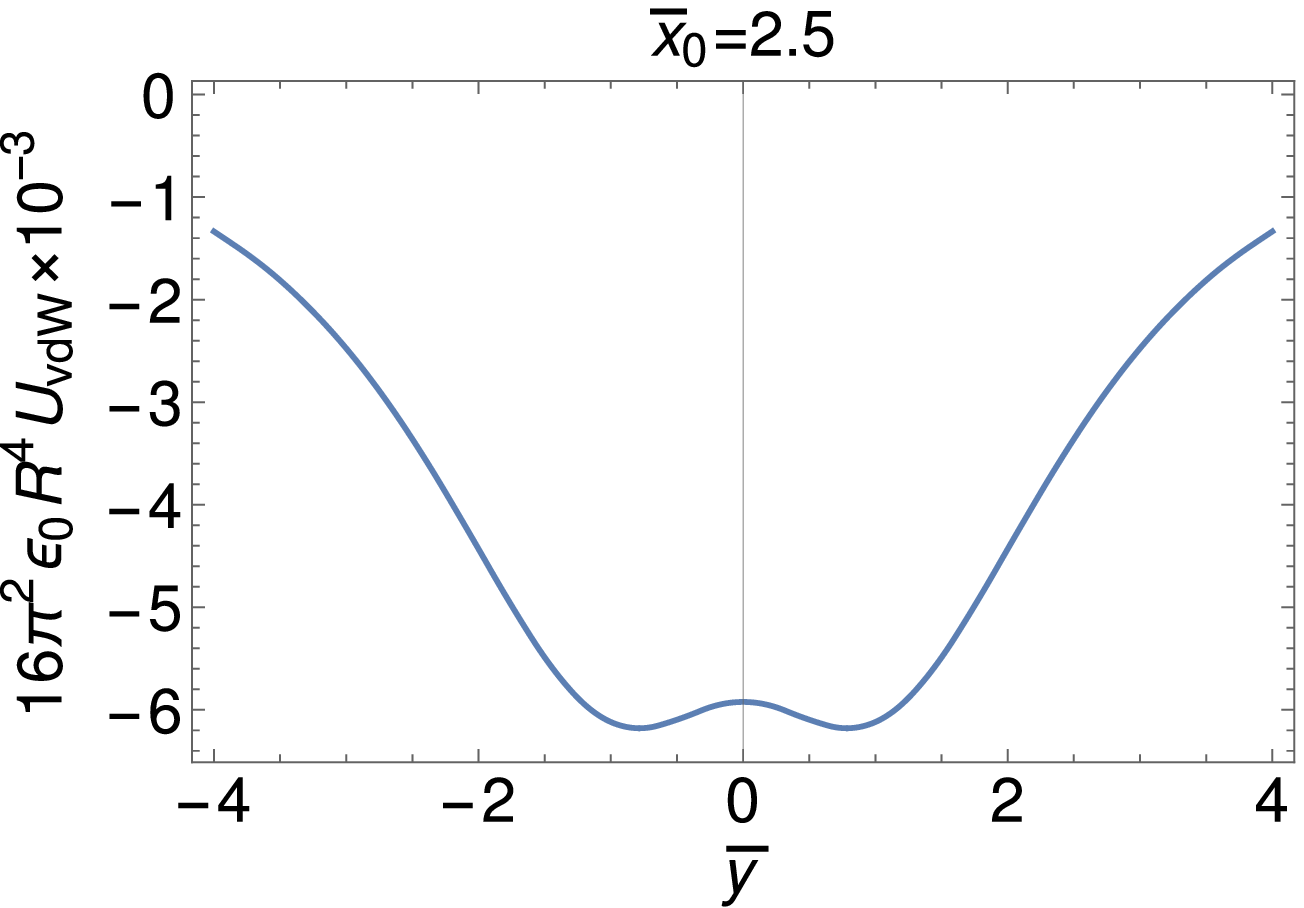, width=0.48 \linewidth}}
\hspace{2mm}
\subfigure[]{\label{fig:vdW3}\epsfig{file=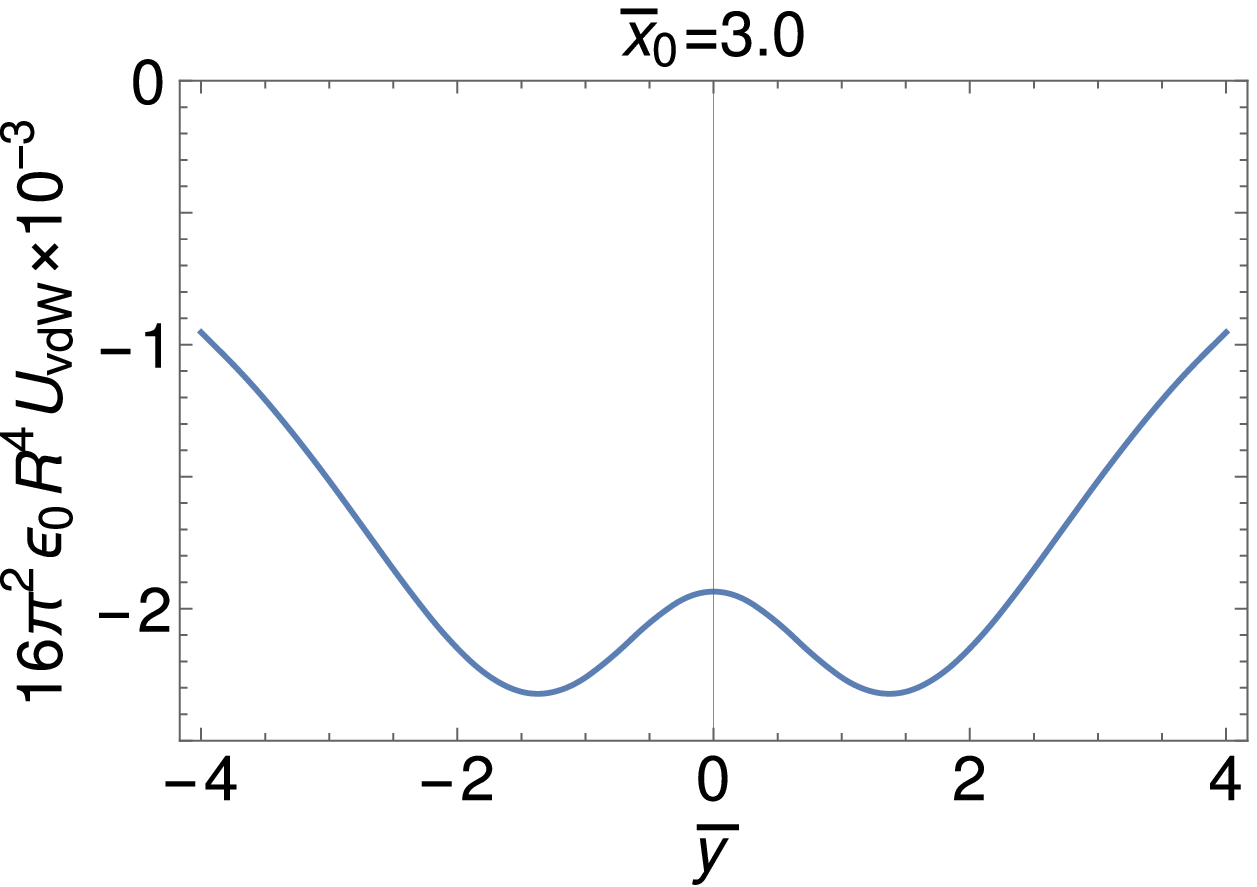, width=0.48 \linewidth}}
\caption{
The behavior of $U_{\text{vdW}}(\overline{x}_0,\overline{y})$
for some values of $\overline{x}_0$.
In (a), $\overline{x}_0=1.50$. 
In (b), $\overline{x}_0=2.18$.
In (c), $\overline{x}_0=2.50$. 
In (d), $\overline{x}_0=3.00$.
Panels (a) and (b) correspond to the dark region in Fig. \ref{fig:brasilia2}, whereas (c) and (d) to the light one.
}
\label{fig:vdW}
\end{figure}

\begin{figure}
	\epsfig{file=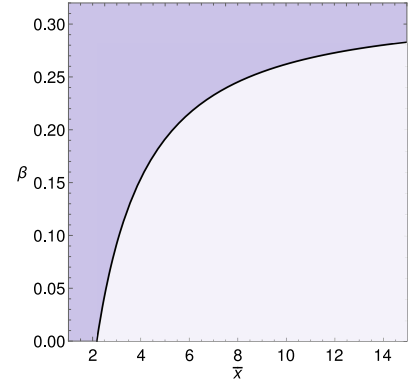,  width=0.6 \linewidth}
	\caption{
		For a particle characterized by 
		$\langle \hat{d}_{x}^2\rangle = \langle \hat{d}_{z}^2\rangle = \beta \langle \hat{d}_{y}^2\rangle$, it is shown the configurations of $\beta$ and $\overline{x}$ for which $y=0$ is a minimum (dark region) or a maximum (light region) point of $U_{\text{vdW}}(x_0,y)$.
		The border between these two regions is given by the curve 
		$\overline{x}=\gamma(\beta)$.
		We remark that $\gamma(0)\approx 2.18$ (the situation discussed in Fig. \ref{fig:brasilia2}), and that above $\beta\approx 0.32$ we just have a dark region.
	}
	\label{fig:regioes-vdW}
\end{figure}

For the vdW case, we carry out an analysis similarly to that done for the CP case. 
We start, again, focusing on the idealized case of a particle characterized by
$\langle \hat{d}_{x}^2\rangle = \langle \hat{d}_{z}^2\rangle = 0$.
For a particle kept constrained to move on a given plane $\overline{x}=\overline{x}_0>1$, from Eq. \eqref{eq:Eberlein-Zietal-vdW}, we have that the dashed and dot-dashed lines shown in Fig. \ref{fig:brasilia2} represent the minimum and maximum points of $U_{\text{vdW}}(x_0,y)$, respectively.
In this figure, the division between the dark and light regions occurs at $\overline{x} \approx 2.18$.
The behavior of ${U}_{\text{vdW}}(\overline{x}_0,\overline{y})$, for some values of $\overline{x}_0$, is shown in Fig. \ref{fig:vdW}.
For a particle with 
$\langle \hat{d}_{x}^2\rangle = \langle \hat{d}_{z}^2\rangle = \beta \langle \hat{d}_{y}^2\rangle$,
we show in Fig. \ref{fig:regioes-vdW} configurations of $\beta$ and $\overline{x}$ for which $y=0$ is a minimum (dark region) or a maximum (light region) point of $U_{\text{vdW}}(x_0,y)$.
For $\beta \gtrsim 0.32$ we just have a dark region, so that any repulsive effect of the lateral force is suppressed.
For $0\leq \beta \leq 0.32$, we can have dark and light regions, depending on $\overline{x}$ and $\beta$.
The border between these two regions is given by the curve 
$\overline{x}=\gamma(\beta)$.
In this figure, note that $\gamma(0)\approx 2.18$, which corresponds to the situation discussed in Fig. \ref{fig:brasilia2}.
In summary, for a particle with $\beta \gtrsim 0.32$ the behavior $F_{\text{vdW}}^{(y)}$ is given by the dark region in Fig. \ref{fig:brasilia2} for all $\overline{x}$, and only particles with $\beta \lesssim 0.32$ can exhibit the repulsive behavior of $F_{\text{vdW}}^{(y)}$ shown by the light region in Fig. \ref{fig:brasilia2}.
Lastly, these curvature-induced effects are affected by the particle orientation 
in relation to the plane where it is kept constrained to move, similar to the CP case, so that if we consider a particle oriented as illustrated in Fig. \ref{fig:brasilia-sem-asa}, we also have a force $F_{\text{vdW}}^{(x)}$ that always take the particle to $x=0$, regardless of the value of $\overline{y}$ (see Fig. \ref{fig:brasilia-sem-asa}).

It is worth to mention that these curvature-induced effects have a classical counterpart.
The interaction energy $U_\text{cla}$, between a perfectly reflecting conducting infinite cylinder and a neutral point particle with a dipole moment vector $\textbf{d}=d_x\hat{{\bf x}}+d_y\hat{{\bf y}}+d_z\hat{{\bf z}}$, is given by replacing $\langle d_{i}^{2}\rangle \to d_{i}^{2}$ in Eq. \eqref{eq:Eberlein-Zietal-vdW} \cite{Eberlein-PRA-2007}.
In this way, for the case of a particle with $d_{x}=d_{z}=0$, we have that the dashed and dot-dashed lines shown in Fig. \ref{fig:brasilia2} also represent the minimum and maximum points of $U_{\text{cla}}(x_0,y)$, respectively.
\begin{figure}
\centering
\epsfig{file=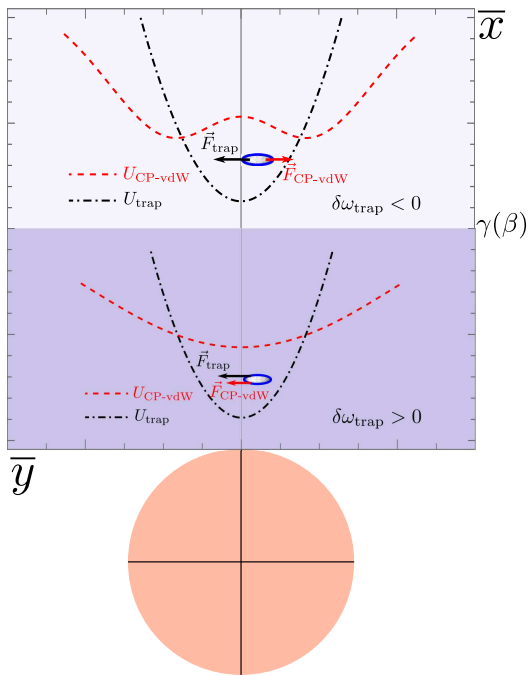,  width=0.8 \linewidth}
\caption{
Illustration of a particle kept constrained to move on a given plane  $\overline{x}=\overline{x}_0>1$,
and subjected to a trap harmonic potential $U_\text{trap}(y)$ (illustrated by the dot-dashed lines) with equilibrium point at $y=0$,	near a perfectly reflecting cylinder. 
The dashed lines represent $U_{\text{CP-vdW}}(x_0,y)$.
The presence of the cylinder modifies this oscillation frequency to a new value $\omega_\text{trap}^\prime$, resulting in a frequency deviation $\delta\omega_\text{trap}\equiv \omega_\text{trap}^\prime-\omega_\text{trap}$.
In dark region [$1 < \overline{x} < \gamma(\beta)$], an experimental apparatus would detect $\delta\omega_\text{trap}>0$.
Otherwise, in light region [$\gamma(\beta) < \overline{x} $], it would detect $\delta\omega_\text{trap}<0$.
}
\label{fig:armadilha}
\end{figure}

\section{Discussions and Final Remarks}
\label{sec:final}

The CP-vdW total force 
${\bf F}_{\text{CP-vdW}}=
-\boldsymbol{\nabla}U_{\text{CP-vdW}}$ 
always attracts the particle to a point closer to the cylinder, so that, focusing
on ${\bf F}_{\text{CP-vdW}}$, the curvature effect 
may not be directly evident.
This effect is a very subtle one, in the sense that it is easier to be perceived when considering the particle kept constrained to move on a given plane near the cylinder.
When the relative curvature becomes sufficiently high, for certain particle orientations and anisotropy, the projected CP-vdW force onto the mentioned plane changes its sign and, instead of moving the particle to the point on the plane closest to the cylinder surface, moves it away.
We considered the existence of such curvature-induced sign inversion effect as our initial research
hypothesis, guided by the sign inversions in the lateral force predicted in
previous works \cite{Nogueira-PRA-2021,Nogueira-PRA-2022}, but,
differently of the models discussed in Refs. \cite{Nogueira-PRA-2021,Nogueira-PRA-2022},
where the infinite extent of the surface did not allow a direct 
characterization of the mentioned sign inversion as a change from an attractive to a repulsive 
effect of the lateral force, for the cylinder this characterization can be done directly.

We showed that the repulsive effect on the lateral force requires a larger cylinder curvature in the retarded (CP) regime than in the nonretarded (vdW) one: the minimum relative curvature required in the CP regime is $6.44$ [see Fig. \ref{fig:regioes-cp}], whereas in vdW is $2.18$ [see Fig. \ref{fig:regioes-vdW}].

The detection of such geometric effects on the lateral CP-vdW force could be done, for example, by trapping an anisotropic particle, and measuring the deviation in the original trap frequency (see, for instance, Refs. \cite{Buhmann-DispersionForces-I, Dalvit-PRL-2008, Nogueira-PRA-2021,Nogueira-PRA-2022}).
Let us consider a particle with mass $m$ and $\alpha_{xx}(0)=\alpha_{zz}(0)=\beta \alpha_{yy}(0)$, with $\beta<0.2$, and the CP regime [or $\langle \hat{d}_{x}^2\rangle = \langle \hat{d}_{z}^2\rangle = \beta \langle \hat{d}_{y}^2\rangle$, with $\beta<0.32$, and the vdW regime], kept constrained to move on a given plane  $\overline{x}=\overline{x}_0>1$, and subjected to a trap harmonic potential $U_\text{trap}(y)$ with equilibrium point at $y=0$ (see Fig. \ref{fig:armadilha}), and oscillating 
with a certain frequency $\omega_\text{trap}$ in the absence of the cylinder.
The presence of the cylinder modifies this oscillation frequency to a new value $\omega_\text{trap}^\prime=\sqrt{\omega_\text{trap}^2+m^{-1}\left[\partial^2U_\text{CP-vdW}(x_0,y)/\partial y^2\right]}$, resulting in a frequency deviation 
$\delta\omega_\text{trap}\equiv \omega_\text{trap}^\prime-\omega_\text{trap}$.
When $1<\overline{x}_0< \gamma(\beta)$ [with $\gamma(\beta)$ given in Figs. \ref{fig:regioes-cp} and \ref{fig:regioes-vdW} for the CP and vdW cases, respectively], an experimental apparatus would detect $\delta\omega_\text{trap}>0$ (see dark region in Fig. \ref{fig:armadilha}).
Otherwise, when $\gamma(\beta)<\overline{x}_0$, one would have $\delta\omega_\text{trap}<0$ (see light region in Fig. \ref{fig:armadilha}).
Therefore, the sign inversion in the lateral force, changing from an attractive behavior to a repulsive one (as illustrated in Fig. \ref{fig:armadilha}), manifests, in this scenario, as a sign inversion in the frequency deviation $\delta\omega_\text{trap}$, with this behavior revealing a nontrivial dependence of the	CP-vdW interaction with	the surface geometry, specifically of the relative curvature.

\begin{acknowledgments}
	L.Q. and E.C.M.N. were supported by the Coordena\c{c}\~{a}o de Aperfei\c{c}oamento de Pessoal de N\'{i}vel Superior - Brasil (CAPES), Finance Code 001.
\end{acknowledgments}
%


%

\end{document}